\documentclass[aps,prl,twocolumn,preprintnumbers,amsmath,amssymb,superscriptaddress]{revtex4-2}
\usepackage{amsmath,graphicx}
\usepackage[utf8]{inputenc}
\usepackage[T1]{fontenc}
\usepackage{xcolor}
\usepackage{textcomp}
\usepackage{bm}

\usepackage{threeparttable}

\usepackage{siunitx}
\usepackage{physics}
\usepackage{amsmath}
\usepackage{tikz}
\usepackage{mathdots}
\usepackage{yhmath}
\usepackage{cancel}
\usepackage{color}
\usepackage{array}
\usepackage{multirow}
\usepackage{amssymb}
\usepackage{gensymb}
\usepackage{tabularx}
\usepackage{extarrows}
\usepackage{booktabs}
\usetikzlibrary{fadings}
\usetikzlibrary{patterns}
\usetikzlibrary{shadows.blur}
\usetikzlibrary{shapes}

\usepackage{color}
\definecolor{LinkColor}{rgb}{0.75,0.0,0.2}

\usepackage{hyperref}
\hypersetup{
	pdfauthor={good guys},
	pdftitle={good title},
	colorlinks=true,
	citecolor=LinkColor,
	linkcolor=LinkColor,
	urlcolor=LinkColor,
}

\usepackage{listings}
\definecolor{lightgray}{gray}{1}

\begin{document}
\title{Supersymmetry dynamics on Rydberg atom arrays}
\author{Shuo Liu}
\thanks{These two authors contributed equally to this work.}
\affiliation{Institute for Advanced Study, Tsinghua University, Beijing 100084, China}

\author{Zhengzhi Wu}
\thanks{These two authors contributed equally to this work.}
\affiliation{Institute for Advanced Study, Tsinghua University, Beijing 100084, China}
\affiliation{Rudolf Peierls Centre for Theoretical Physics, Parks Road, Oxford, OX1 3PU, UK}

\author{Shi-Xin Zhang}
\email{shixinzhang@iphy.ac.cn}
\affiliation{Institute of Physics, Chinese Academy of Sciences, Beijing 100190, China}

\author{Hong Yao}
\email{yaohong@tsinghua.edu.cn}
\affiliation{Institute for Advanced Study, Tsinghua University, Beijing 100084, China}

\date{\today}

\begin{abstract}
Spacetime supersymmetry (SUSY) that interchanges fermions and bosons is of great theoretical importance but has not yet been revealed experimentally in particle physics. It has also been desired to explore quantum-mechanical SUSY in microscopic lattice models. Inspired by the recent experiments of Floquet engineering of Rydberg atom arrays, we propose to simulate quantum mechanical supersymmetric model and realize quantum mechanical SUSY in Floquet Rydberg atom arrays. Moreover, we utilize the supercharge dynamics to demonstrate the SUSY property of the model under investigation: the expectation value of supercharge freezes under time evolution for supersymmetric lattice models in contrast to the trivial oscillation for generic nonsupersymmetric lattice models. The proposal is validated on direct simulation of Rydberg atom arrays' dynamics and ready for experiments. This work sheds light on the future experimental exploration of SUSY with the help of Rydberg atom arrays.
\end{abstract}

\maketitle

\textit{Introduction.---} The concept of spacetime supersymmetry (SUSY)~\cite{GERVAIS1971632, WESS197439}, which postulates a symmetry interchanging fermions and bosons, offers a compelling way to address various fundamental and long-standing problems, including the hierarchy problem in particle physics ~\cite{DIMOPOULOS1981150} and cosmological constant problem~\cite{CREMMER198361}. However, despite extensive research efforts, there has been no conclusive experimental evidence to date for SUSY and its spontaneous breaking within the realm of particle physics. 

On a different front, it is well known that spacetime SUSY might emerge at low energies in many-body systems at criticality. For instance, the spacetime SUSY could emerge at 1+1 dimensional tricritical Ising transition~\cite{FRIEDAN198537, Zamolodchikov:1986db}. Emergent SUSY at quantum
criticality and the associated microscopic models have been extensively investigated~\cite{PhysRevLett.52.1575, doi:10.1142/S0217979298000570, 
PhysRevB.76.075103, PhysRevLett.100.090404,
PhysRevB.80.075432,
PhysRevLett.105.150605,
RevModPhys.82.3045,
RevModPhys.83.1057,
PhysRevB.87.165145,
PhysRevB.87.041401,
doi:10.1126/science.1248253,
Ponte_2014,
PhysRevLett.114.090404,
PhysRevLett.114.237001,
PhysRevLett.115.166401,
PhysRevLett.117.166802, 
PhysRevLett.118.166802,
PhysRevLett.119.107202,
doi:10.1126/sciadv.aau1463,
PhysRevLett.120.206403,PhysRevLett.126.206801, li2024uncovering}. 
Even so, the experimental confirmation of the emergent spacetime SUSY remains elusive. On one hand, investigating the ground state properties presents a significant challenge for existing experimental platforms, particularly when attempting to achieve emergent SUSY at a multicritical point, which requires fine-tuning of more than one parameter and adiabatic annealing. On the other hand,  to accurately characterize SUSY, it is essential to employ unconventional measurement methods \cite{doi:10.1126/science.1209284, doi:10.1126/science.aam8990, doi:10.1126/science.aav3587, doi:10.1126/science.aav9105} capable of detecting the non-local string operators that are intrinsic to fermionic modes.

Apart from lattice models with emergent spacetime SUSY, lattice models with explicit quantum-mechanical SUSY can be constructed~\cite{PhysRevLett.90.120402, PaulFendley_2003,XiaoYang_2004, PhysRevLett.95.046403, PhysRevLett.101.146406, PhysRevB.84.115124, Huijse_2012,PhysRevB.100.195146,PhysRevLett.126.236802,PhysRevB.103.085130,cai_observation_2022,PhysRevB.110.165124}. One well-known family of supersymmetric lattice models is the $M_{k}$ model~\cite{PaulFendley_2003} which describes 1D fermion chain under the constraint that at most $k$ consecutive sites may be occupied and thus restricting the hopping. Explicit quantum-mechanical SUSY of a lattice model 
can induce a nontrivial energy spectrum -- all its eigenstates form either singlet or doublet representation of the SUSY algebra. 
The singlet corresponds to the supersymmetric ground state with eigenenergy $E=0$, while each doublet includes two eigenstates with the same eigenenergy $E>0$. The fermion numbers of two eigenstates in each doublet differ by 1 and thus these two eigenstates correspond to fermionic and bosonic modes, respectively. Consequently, the supersymmetric lattice model also exhibits unique dynamical features which could be more experimental accessible than the ground state properties in a variety of quantum simulator platforms. 

Rydberg atom arrays serve as a powerful and flexible platform for processing quantum information~\cite{RevModPhys.82.2313, 10.1116/5.0036562, bluvsteinQuantumProcessorBased2022a, maHighfidelityGatesMidcircuit2023, schollErasureConversionHighfidelity2023, doi:10.1126/science.ade5337, bluvsteinLogicalQuantumProcessor2024a, xuConstantoverheadFaulttolerantQuantum2024},
exploring exotic phases of matter and simulating novel quantum dynamics~\cite{bernien_probing_2017, browaeys_many-body_2020, ebadi_quantum_2021,scholl_quantum_2021, Wu_2021, doi:10.1126/science.abg2530, doi:10.1126/science.abi8794,fang2024probing, cheng_emergent_2024,  wuDissipativeTimeCrystal2024, anomalousinformationscrambling}. Recently, a Floquet engineering scheme of blockade-consistent spin-exchange
interactions in periodically driven Rydberg atom arrays has been proposed~\cite{Floquet}, naturally resembling the constraint in supersymmetric $M_{k}$ models~\cite{PaulFendley_2003} and indicating the potential of utilizing Rydberg atom arrays to construct non-trivial interactions in quantum many-body systems. With the Floquet engineering toolbox on Rydberg atom arrays, it is important and urgent to explore the new possibility of experimentally demonstrating SUSY.

\begin{figure*}[t]
\centering
\includegraphics[width=0.98\textwidth, keepaspectratio]{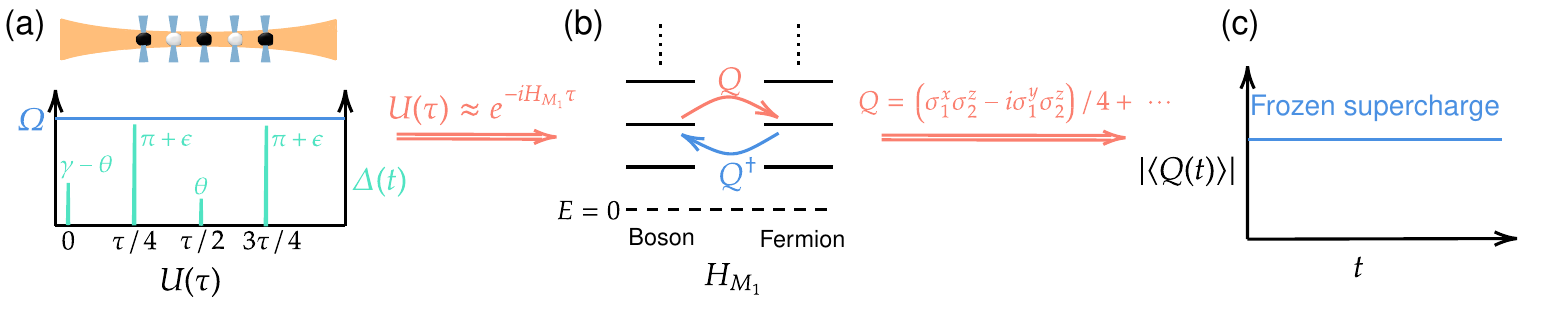}
\caption{Simulating supersymmetry dynamics on Rydberg atom arrays. (a) Upper panel: Rydberg atom arrays within optical tweezers. The degree of freedom at each site comprises the atomic ground state ($\vert g \rangle$, black solid circle) and the Rydberg state (highly excited state $\vert r \rangle$, white solid circle). Lower panel: the time evolution within one period $\tau$ and the unitary evolution operator over one period is given by $U(\tau)$ (see Eq.~\eqref{eq:Utau_main}). Firstly, we utilize the PXP model as an approximate description of Rydberg atoms and the static Hamiltonian is $    H^{\text{obc}}_{\text{PXP}} =\sigma^{x}_{1}P^{r}_{2} + \sum_{i=2}^{L-1} P^{r}_{i-1}\sigma^{x}_{i} P^{r}_{i+1} + P^{r}_{L-1} \sigma^{x}_{L}$ whose strength is denoted as Rabi frequency $\Omega$ (blue line). The four green peaks represent the global detunings at times $0$, $\tau/4$, $\tau/2$, $3\tau/4$, respectively. The strength of the global detuning, i.e., the height of the peak, is parameterized by $\gamma$, $\theta$, $\epsilon$ and tunable. (b) Secondly, by employing suitable Floquet engineering as discussed in the main text, we achieve a Floquet effective Hamiltonian that mirrors the supersymmetric $M_{1}$ model. It is a supersymmetric lattice model and the eigenstates with $E>0$ naturally form doublets, giving two degenerate states with different fermion parity. (c) Lastly, we employ the expectation value of the supercharge operator to investigate the supersymmetry dynamics. Notably, the dynamics of the supercharge operator is a constant under the quench of a supersymmetric Hamiltonian.}
\label{fig:schematic}
\end{figure*}

In this Letter, we find that quantum-mechanical SUSY can be realized in Floquet Rydberg atom arrays by tuning pulse parameters, inspired by the recent Floquet engineering of Rydberg atom arrays~\cite{Floquet}. We further propose to experimentally investigate the supersymmetry dynamics under the Floquet engineering~\cite{PhysRevLett.20.180, PhysRevLett.111.185301, PhysRevLett.111.185302, PhysRevX.4.031027, 
jotzuExperimentalRealizationTopological2014,
doi:10.1080/00018732.2015.1055918, 
PhysRevLett.116.205301, 
doi:10.1126/science.aad4568, RevModPhys.89.011004,  
PhysRevLett.120.070501, 
PhysRevX.10.031002, doi:10.1126/science.abd9547, PRXQuantum.3.020303, PhysRevLett.130.120403,PhysRevLett.131.220803, geier2024timereversaldipolarquantummanybody, Floquet}. Specifically, we demonstrate that the effective Floquet Hamiltonian of the Rydberg atom array can be tuned to the supersymmetric $M_{1}$ model \cite{PhysRevLett.90.120402} on an open-boundary chain, which is the simplest one in the family of supersymmetric $M_{k}$ models. Moreover, we propose to characterize the SUSY dynamics by the evolution of the supercharge which has friendly experimental requirements in terms of state preparation and measurements.
The supersymmetric Hamiltonian with the frozen supercharge dynamics can be distinguished from other nonsupersymmetric Hamiltonians where the supercharge dynamics is nonconserved. We have also numerically demonstrated the non-trivial supersymmetry dynamics using both effective Hamiltonian and the original Rydberg atom array dynamics and analyzed the effects of nonsupersymmetric perturbations. The main results are summarized in Fig.~\ref{fig:schematic}. Besides supercharge dynamics, this scheme can be directly applied to investigate the dynamics of the supersymmetric lattice model characterized by other observables. This provides a pathway to explore other intriguing dynamics related to SUSY such as the weak ergodicity breaking from supersymmetry~\cite{buijsman2024weak}.

\textit{The $M_{1}$ model and supersymmetry dynamics.---}We consider the one-dimensional $M_{1}$ model~\cite{PhysRevLett.90.120402}, which is one of the most famous $\mathcal{N}=2$ supersymmetric lattice model, with the following Hamiltonian 
\begin{eqnarray}
    H_{M_{1}} = \{Q, Q^{\dagger} \}.\label{eq:m1}
\end{eqnarray}
$Q = \sum_{i} Q_{i}$ is the nilpotent supercharge satisfying $Q^2=0$, where $Q_{i} = (-1)^{i} P_{i-1} c_{i}^{\dagger} P_{i+1}$ with $c_{i}^{\dagger}$ and $P_{i} \equiv 1-c^{\dagger}_{i} c_{i}$ being the fermion creation operator and projection operator to vacuum state at site $i$. Obviously, $[Q, H_{M_{1}}] = [Q^{\dagger}, H_{M_{1}}]=0$; namely the lattice Hamiltonian in Eq. \eqref{eq:m1} explicitly processes the supersymmetry. Moreover, based on the construction of Hamiltonian, the eigenvalues are non-negative and thus an eigenstate $\vert G \rangle$ with $E=0$ must be a ground state, which is dubbed as supersymmetric ground state satisfying $Q\vert G \rangle = Q^{\dagger} \vert G \rangle= 0$. Other eigenstates with $E >0$ form doublet representations. Each doublet includes two eigenstates: $\vert s \rangle$ and $Q \vert s \rangle$, satisfying $Q^\dagger \vert s \rangle =0$ and $Q (Q\vert s \rangle)=0$. Moreover, $Q\vert s \rangle $ is the superpartner of $\vert s \rangle$ sharing the same eigenenergy yet exhibiting a distinct fermion parity as supercharge operator $Q$ effectively adds a fermion into the system. Consequently, two eigenstates in each doublet can be regarded as the fermionic and bosonic modes, respectively, which is a hallmark of quantum-mechanical SUSY. The number of supersymmetric ground states depends on the choice of system size and boundary conditions (see Ref.~\cite{PaulFendley_2003} for detailed discussion). In the following, we focus on the case with system size $L=3l+1$ and open boundary conditions (OBC). There is no supersymmetric ground state, nevertheless, all the eigenstates can still be decomposed into doublets under SUSY.  

Owing to the presence of explicit quantum-mechanical SUSY in the $M_1$ model, we aim to extract nontrivial features unique to SUSY from quantum dynamics. 
The experimental demonstration of SUSY dynamics will provide a practical and meaningful task for current quantum devices to achieve an advantage beyond the reach of the classical simulation. 
For instance, the quench dynamics for a generic Hamiltonian vary significantly depending on initial states while the dynamics starting from a certain initial state are the same as those starting from its superpartner guaranteed by the supersymmetry~\cite{PhysRevLett.128.050504}. Therefore, the observation of identical quantum dynamics from two initial states serves as compelling evidence for the existence of SUSY. However, akin to the unrealized experimental exploration of spacetime SUSY in low energy, simulating SUSY dynamics poses a significant challenge due to the difficulty in preparing the superpartner of a generic state and thus the experimental investigations are severely limited (see Supplemental Materials (SM) for more details~\footnote{see Supplemental Materials for more details, including (I) advantage of demonstrating supersymmetry via supercharge dynamics, (II) comparison with other experimental proposals for SUSY on Rydberg atom arrays, (III) details of initial state preparation and supercharge measurement, (IV) derivation of $M_{1}$ model in the spin-1/2 basis, (V) derivation of the effective Floquet Hamiltonian, (VI) details of numerical simulation and additional numerical results.}).

To address this limitation, we propose to directly investigate the dynamics of the expectation value of supercharge operator $Q$ 
\begin{eqnarray}
    \langle Q(t) \rangle = \langle \psi(0) \vert e^{iHt} Q e^{-iHt} \vert \psi(0) \rangle,
\end{eqnarray}
to reveal the supersymmetry, where $\vert \psi(0) \rangle $ is an abitrary initial state and $H$ is a generic Hamiltonian. When $H$ is supersymmetric, for instance $H=H_{M_{1}}$ and $[Q, H]=0$, the supercharge dynamics freeze, 
\begin{eqnarray}
     \langle Q(t) \rangle = \langle \psi(0) \vert Q \vert \psi(0) \rangle = \langle Q(0) \rangle =c,
\end{eqnarray}
where $c$ is an initial-state dependent (complex) constant. In contrast, $\langle Q(t) \rangle$ will be nonconserved for a generic Hamiltonian that does not  
respect SUSY. Consequently, the frozen supercharge dynamics serve as a decisive signal for the presence of SUSY, and no high-precision initial state preparation is required. 
More importantly, as detailed below, both the analog simulation of the $M_{1}$ model and the estimation of the expectation value of the supercharge operator can be easily realized on Rydberg atom arrays.

In the following, we shall focus on the $M_{1}$ model with OBC denoted as $H_{M_{1}}^{\text{obc}}$. With the help of the Jordan-Wigner transformation and particle-hole transformation, the $M_{1}$ model in the spin-1/2 basis can be rewritten as
\begin{eqnarray}
    \label{eq:HM1_main}
    H^{\text{obc}}_{M_{1}} &=& \frac{1}{4} H^{\text{obc}}_{\text{ZIZ}} - H^{\text{obc}}_{\text{PXYP}} - N \\ \nonumber
    &+& \frac{1}{2}(n^{r}_{1}-n^{r}_{2}-n^{r}_{L-1} + n^{r}_{L}),
\end{eqnarray}
where $H^{\text{obc}}_{\text{ZIZ}}$ is the next-nearest-neighbor $ZZ$ interaction $H^{\text{obc}}_{\text{ZIZ}} = \sum_{j=2}^{L-1} \sigma^{z}_{j-1} \sigma^{z}_{j+1}$, $H^{\text{obc}}_{\text{PXYP}}$ is the blockaded nearest-neighbor spin-exchange interaction $H^{\text{obc}}_{\text{PXYP}} = \frac{1}{2} (\sigma^{x}_{1}\sigma^{x}_{2} + \sigma^{y}_{1}\sigma^{y}_{2})P^{r}_{3}
+ \frac{1}{2}\sum_{j=2}^{L-2} P^{r}_{j-1} (\sigma^{x}_{j}\sigma^{x}_{j+1} + \sigma^{y}_{j}\sigma^{y}_{j+1})P^{r}_{j+2} + \frac{1}{2}P^{r}_{L-2}(\sigma^{x}_{L-1}\sigma^{x}_{L} + \sigma^{y}_{L-1}\sigma^{y}_{L})$, and $N$ is the number operator $N =\sum_{j=1}^{L} n_{j}^{r}$  with $P_{i}^{r} = \frac{\mathbb{I} + \sigma^{z}_{i}}{2}$ and $n_{i}^{r} = \frac{\mathbb{I} - \sigma^{z}_{i}}{2}$. Here $\sigma_{i}^{(x,y,z)}$ are Pauli matrix on site $i$. As shown below, the local spin-1/2 Hilbert space can be formed by the Rydberg atom's analog basis $\vert g \rangle$ and $\vert r \rangle$, representing ground and Rydberg states, respectively. $P_{i}^{r}=\vert g \rangle \langle g \vert_{i}$ is the projection operator to the ground state $\vert g \rangle$ on site $i$ while $n_i^r=\vert r \rangle \langle r \vert_{i}$ projects to the Rydberg state. 
Moreover, the supercharge operator $Q$ corresponds to a summation of Pauli string operators: $Q = \frac{\sigma^{-}_{1}}{2} \frac{\mathbb{I}+\sigma^{z}_{2}}{2} + \sum_{i=2}^{L-1} \frac{\mathbb{I}+\sigma^{z}_{i-1}}{2} \left( \frac{\sigma^{-}_{i}}{2} \prod_{j=1}^{i-1} (-\sigma^{z}_{j}) \right) \frac{\mathbb{I}+\sigma^{z}_{i+1}}{2} + \frac{\mathbb{I}+\sigma^{z}_{L-1}}{2} \left(\frac{\sigma^{-}_{L}}{2} \prod_{j=1}^{L-1} (-\sigma_{j}^{z}) \right)$. See SM for more details~\cite{Note1}.

\textit{Experimental implementation.---}Recently, Floquet engineering of blockaded spin-exchange interaction in periodically driven Rydberg chains has been proposed and utilized for the exploration of gapless Luttinger liquid phase~\cite{Floquet}. Inspired by this construction, we harness the potential of Floquet engineering and propose to experimentally simulate the supercharge dynamics of the supersymmetric $M_{1}$ model on Rydberg atom arrays.

For simplicity, we consider the following driven PXP model with OBC 
\begin{eqnarray}
    \label{eq:Hpxp_main}
    H^{\text{obc}}(t) = \frac{\Omega}{2}H^{\text{obc}}_{\text{PXP}}-\Delta_{0}(t) N - \tilde{\Delta}(t) N,
\end{eqnarray}
where 
\begin{eqnarray}
    H^{\text{obc}}_{\text{PXP}} &=& \sigma^{x}_{1}P^{r}_{2} + \sum_{i=2}^{L-1} P^{r}_{i-1}\sigma^{x}_{i} P^{r}_{i+1} + P^{r}_{L-1} \sigma^{x}_{L},
\end{eqnarray}
is the pure PXP Hamiltonian, which describes the blockade mechanism in a chain of Rydberg atoms and the effect of sandwiching the Pauli $X$ operators on each site between projectors is that the state of an atom can flip between the ground state and Rydberg state only if both adjacent atoms are in the
ground states.
Here, $\Delta_{0}(t)$ and $\tilde{\Delta}(t)$ are the time-dependent global detunings given by
\begin{eqnarray}
    \Delta_{0}(t) &=& \pi \sum_{n} \delta(t-\frac{\tau}{4}-n\frac{\tau}{2}), \\
        \Tilde{\Delta}(t) &=& (\gamma-\theta) \sum_{n} \delta(t-n\tau) + \theta \sum_{n} \delta(t-\frac{\tau}{2}-n\tau) \\ \nonumber
        &+& \epsilon \sum_{n} \delta(t-\frac{\tau}{4}-n\tau) + \epsilon \sum_{n} \delta(t-\frac{3\tau}{4}-n\tau),  
\end{eqnarray}
with $\tau$ being the Floquet period and $\gamma$, $\theta$, $\epsilon$ being the controllable strengths of global detuning perturbations. The corresponding time evolution unitary for one period is 
\begin{eqnarray}
    \label{eq:Utau_main}
    U(\tau) &=& \mathcal{T} e^{-i \int_{0}^{\tau} dt H^{\text{obc}}(t)} \\ \nonumber
   &=& U_{\text{PXP}} V_{N}(\epsilon+\pi) U_{\text{PXP}} V_{N}(\theta)  \\ \nonumber
    && U_{\text{PXP}} V_{N}(\epsilon+\pi) U_{\text{PXP}} V_{N}(\gamma-\theta),    
\end{eqnarray}
where
\begin{eqnarray}
    U_{\text{PXP}} = e^{-i\frac{\tau}{4}\frac{\Omega}{2} H_{\text{PXP}}^{\text{obc}}}, \ \ \ V_{N}( \beta) &=& e^{i \beta N}.
\end{eqnarray}
It can be shown that Floquet unitary $U_{F}\equiv U(\tau)$ can be approximately described by a time-independent effective Floquet Hamiltonian $H_{F}$, i.e.,
\begin{eqnarray}
    U_F\approx e^{-i H_{F}\tau}, 
\end{eqnarray}
and thus the stroboscopic dynamics ($t=n\tau$, $n \in \mathbb{N}$) corresponds to the time evolution under $H_{F}$: $U(n\tau) = (U_{F})^{n} \approx e^{-i H_{F} n\tau}$. With small detuning perturbations ($\epsilon, \gamma, \theta$), one can obtain the effective Floquet Hamiltonian $H_{F}$ through a Floquet-Magnus expansion~\cite{PhysRevLett.127.090602, PhysRevX.7.011026}: $H_{F} = H_{F}^{(0)}+H_{F}^{(1)}+\mathcal{O}(\epsilon^{3})$.

\begin{figure}[t]
\centering
\includegraphics[width=0.49\textwidth, keepaspectratio]{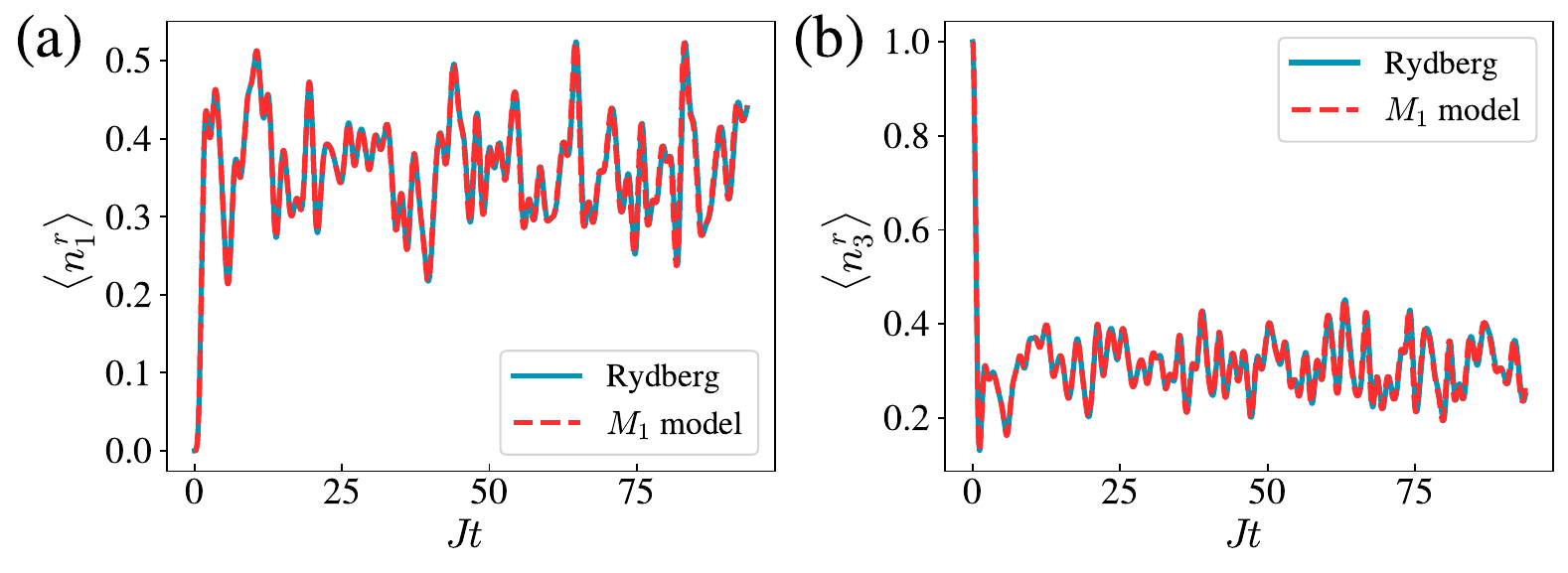}
\caption{The dynamics of Rydberg state density on site 1 (a) and site 3 (b), respectively. Here, we set $L=13$, $\tau=0.001$, $\frac{\Omega\tau}{4}=0.025$, $\epsilon=-0.1$. The numerical results evolved by the $M_{1}$ model (red) agree well with those evolved by the Floquet unitary $U_{F}$ (blue).}
\label{fig:z_dynamics_l4_Omegatau0.1_epsilon0.1}
\end{figure}

When $\frac{\Omega \tau}{4} \ll 1$, 
\begin{eqnarray}
    H_{F}^{(0)} &=& \frac{h}{4} H^{\text{obc}}_{\text{ZIZ}}  -h H_{\text{PXYP}}^{\text{obc}} -J N, \\ \nonumber
    &+&  \frac{h}{2}(n^{r}_{1}-n^{r}_{2}-n^{r}_{N-1}+n^{r}_{N}) \\
    H_{F}^{(1)} &=& g H_{\text{PXP}}^{\text{obc}},
\end{eqnarray}

where
\begin{eqnarray}
    J &=& \frac{\gamma + 2\epsilon}{\tau} - \frac{3\epsilon \Omega^2 \tau}{32}, \\ 
    h &=& -\frac{\epsilon \Omega^2 \tau}{32}, 
    \\ 
    g &=& -\frac{\epsilon(\epsilon+\theta)\Omega}{8}.
\end{eqnarray}
See the SM~\cite{Note1} for more details and Ref.~\cite{Floquet} for the derivation with periodic boundary conditions. Furthermore, when $\epsilon/\Omega \tau \ll 1$, the contribution of the $H_{F}^{(1)}$ becomes negligible. In this case, it is permissible to set $\theta$ to zero, thereby considering only the three global detuning perturbations that occur over each period. More importantly, when  $\gamma = \epsilon (\frac{\Omega^2 \tau^2}{16} - 2)$, i.e. $J=h$, we have
\begin{eqnarray}
    \label{eq:HF_J_main}
    H_{F} \approx H_{F}^{(0)} &=&  \frac{J}{4} H^{\text{obc}}_{\text{ZIZ}} -J H^{\text{obc}}_{\text{PXYP}} -J N  \\ \nonumber
    &+& \frac{J}{2}(n^{r}_{1}-n^{r}_{2}-n^{r}_{N-1}+n^{r}_{N})),
\end{eqnarray}
and thus the zeroth-order $H_{F}$ is exactly the same as the Hamiltonian of the supersymmetric $M_{1}$ model as shown in Eq.~\eqref{eq:HM1_main} up to an unimportant overall factor $J$. Therefore, the driven PXP model under Floquet engineering with suitable parameters can describe the quantum dynamics of the supersymmetric $M_{1}$ model. We note that the first-order correction term $H_{F}^{(1)}$ does not preserve the quantum-mechanical SUSY property. However, it is possible to tune $\theta = -\epsilon$ to eliminate the $H_{\text{PXP}}$ term, thereby ensuring that the effective Floquet Hamiltonian retains SUSY up to this order. In the following, we set $\gamma = \epsilon (\frac{\Omega^2 \tau^2}{16} - 2)$ and $\theta = -\epsilon$ unless other specified. 

To validate the correspondence between the stroboscopic dynamics of the driven PXP model under Floquet engineering (Eq.~\eqref{eq:Utau_main}) and the supersymmetric $M_{1}$ model, we use the python package {\sf TensorCircuit}~\cite{*[{ }] [{. \url{https://github.com/tensorcircuit/tensorcircuit-ng}.}] Zhang2023tensorcircuit} to perform numerical simulations. We calculate the density dynamics of Rydberg states on different sites. The system size is chosen as $L=3l+1$ and the initial state is chosen as 
\begin{eqnarray}
    \label{eq:K1_main}
    \vert K_{1} \rangle = (\prod_{j=1}^{l} \sigma^{x}_{3j} )\vert g \rangle^{\otimes L}.
\end{eqnarray}
As illustrated in Fig.~\ref{fig:z_dynamics_l4_Omegatau0.1_epsilon0.1}, the Rydberg state density dynamics evolved by Floquet unitary $U_{F}$ coincide with those evolved by the $M_{1}$ model. In addition, we have also calculated the dynamics of the correlation function and tested other initial states (see more numerical results in the SM~\cite{Note1}). The results also agree well with those of the $M_{1}$ model. Therefore, the effective Floquet Hamiltonian is a good approximation of the $M_{1}$ model. We note that the accumulated approximation errors will induce the deviation at late times.

\textit{Supercharge dynamics.---}Having demonstrated the effectiveness of simulating the real-time evolution of $M_{1}$ model by the driven PXP model tuned to the SUSY point, we then investigate the supersymmetry dynamics characterized by the expectation value of supercharge $\langle Q(t) \rangle$. 

We note that the initial state should be easily prepared on Rydberg atom arrays, yielding a superposition of eigenstates mixing different fermion-parity sectors to ensure a non-zero $\langle Q(t) \rangle$. To prepare the initial state, we first let the state $\vert K_{1} \rangle$ (see Eq.~\eqref{eq:K1_main}, or any other easily prepared state) evolve with pure PXP Hamiltonian $H_{\text{PXP}}^{\text{obc}}$ breaking the parity symmetry to time $t^{*}$. We choose 
\begin{eqnarray}
\vert \psi(0) \rangle = e^{-iH_{\text{PXP}}^{\text{obc}}t^{*}} \vert K_{1} \rangle
\end{eqnarray}
as the initial state for the supersymmetry dynamics and turn on the Floquet engineering at $t^*$.
Then $\vert \psi(0)\rangle$ is evolved by the Floquet unitary $U_{F}$ and thus $\vert \psi(n\tau) \rangle = (U_{F})^{n} \vert \psi(0) \rangle$.

\begin{figure}[t]
\centering
\includegraphics[width=0.49\textwidth, keepaspectratio]{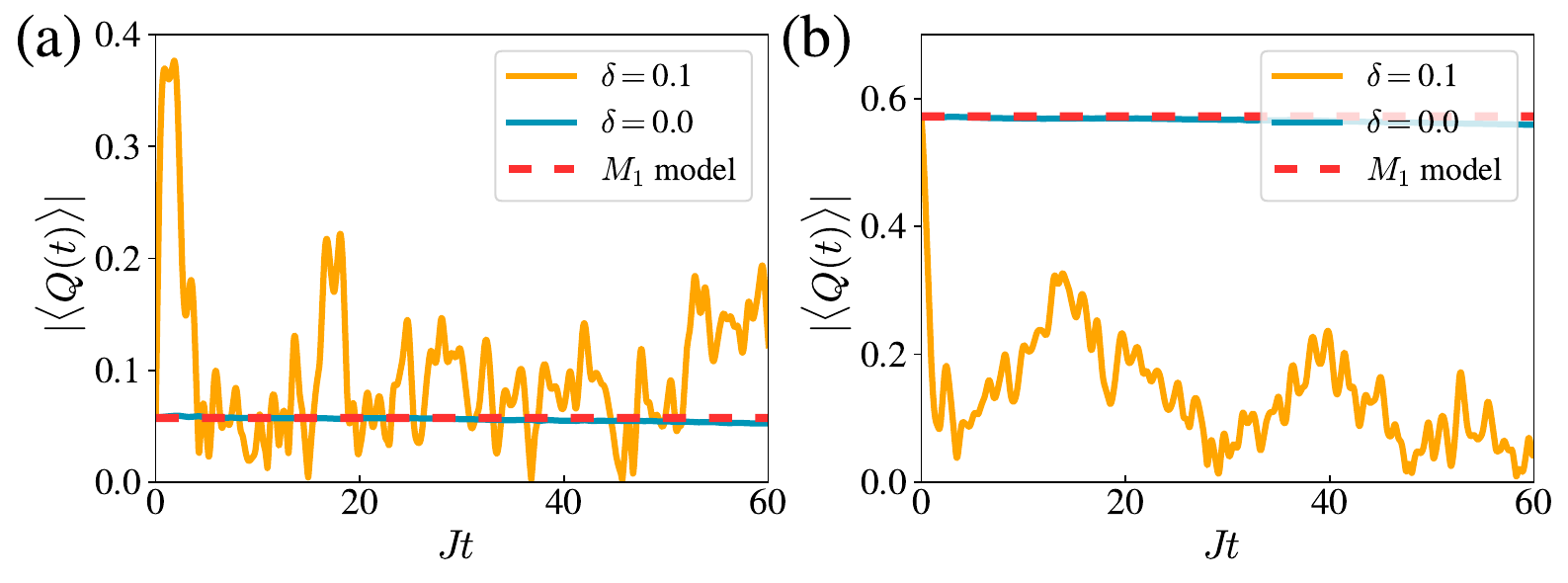}
\caption{Supercharge dynamics with (a) $t^{*}=5$ and (b) $t^{*}=10$. Here, we set $L=13$, $\tau=0.001$, $\frac{\Omega\tau}{4}=0.025$, $\epsilon=-0.1$, and $\theta = -(1+\delta) \epsilon$. When $\delta=0.0$ the expectation value of supercharge is a constant and is the same as that under the quench of supersymmetric $M_{1}$ model, indicating the effective Floquet Hamiltonian feature a quantum-mechanical SUSY. However, when $\delta>0$ the supercharge dynamics exhibit oscillation, i.e., the effective Floquet Hamiltonian is nonsupersymmetric. We note that the accumulated Floquet engineering approximation errors induce the deviation at late times even for $\delta=0$.}
\label{fig:Rydberg_Q_l4_Omegatau0.1_epsilon0.1}
\end{figure}

As illustrated in Eq.~\eqref{eq:HF_J_main}, the effective Floquet Hamiltonian $H_{F}$ is supersymmetric when $\theta = -\epsilon$ and $\gamma = \epsilon (\frac{\Omega^2 \tau^2}{16} - 2)$. Here, we set $\theta = -(1+\delta) \epsilon$ and use $\delta$ to describe the deviation away from the supersymmetric point. 
The numerical results of supercharge dynamics with different initial states are shown in Fig.~\ref{fig:Rydberg_Q_l4_Omegatau0.1_epsilon0.1}. 
When $\delta=0$, the effective Floquet Hamiltonian is the same as the $M_{1}$ Hamiltonian and the expectation value of the supercharge remains unchanged. 
In contrast, when $\delta>0$, the effective Floquet Hamiltonian is not supersymmetric anymore and then there exist decays and oscillations in supercharge dynamics.

Furthermore, to experimentally demonstrate the supersymmetry dynamics of the $M_{1}$ model, two issues must be addressed. 
Firstly, in the experiment, the Rydberg atoms actually interact via van der Waals interaction with blockade radius $R_{b}$ and the global detuning pulses have finite widths. 
Previous work~\cite{Floquet} has validated the good agreement between the results of the driven PXP model and the model of driven Rydberg atom arrays with long-range interactions. 
Therefore, we believe that the corresponding effective Floquet Hamiltonian in the real experiment platform can still act as a good approximation of the supersymmetric $M_{1}$ model. 
Secondly, the measurement of supercharge should be friendly for the Rydberg atom arrays. 
Via the quantum analog-to-digital converter~\cite{bluvsteinQuantumProcessorBased2022a, RevModPhys.89.015006}, one can transform from the analog basis to the digital basis where the measurement of any Pauli string is easy via proper single qubit rotations and computational basis measurements.

\textit{Discussion and outlook.---} In conclusion, we have revealed that the supersymmetric $M_{1}$ model can be realized on Rydberg atom arrays via Floquet engineering by choosing appropriate parameters. We further proposed to uncover the supersymmetry dynamics via the supercharge expectation value and have demonstrated the frozen supercharge dynamics for supersymmetric Hamiltonian. This scheme is suitable for experimental demonstration on current Rydberg atom arrays. 

It would be intriguing to explore the Floquet engineering scheme to achieve $M_{1}$ model with higher precision.
Besides the supersymmetry dynamics of the one-dimensional $M_{1}$ model, it is worth further investigating the quantum simulation in higher dimensions or for other supersymmetric lattice models. More importantly, the current Floquet engineering scheme uses only global detunings. It is  
essential to extend Floquet engineering to the case with site-dependent detuning fields~\cite{manovitz2024quantumcoarseningcollectivedynamics}, which is helpful for constructing other nontrivial interactions. 
Furthermore, due to the flexibility of the Floquet engineering, it is promising to prepare the ground state of the supersymmetric $M_{1}$ model via adiabatic evolution, and thus experimentally identify the spacetime SUSY in low energy regime of the $M_1$ model and other related models~\cite{alcaraz1999exactly}, which shall provide significant light into the understanding of spacetime SUSY and its breaking. 

\textit{Acknowledgements.---} We acknowledge Cheng Chen, Shao-Kai Jian, Zi-Xiang Li, and Shuai Yin, and especially Paul Fendley for helpful discussions. This work was supported in part by NSFC under Grants No. 12347107 and No.12334003 (S.L. and H.Y.), by MOSTC under Grant No.2021YFA1400100 (H.Y.), and by the Xplorer Prize through the New Cornerstone Science Foundation (H.Y.). S.-X.Z. acknowledges the support from Innovation Program for Quantum Science and Technology (2024ZD0301700) and the start-up grant at IOP-CAS. Z.W. acknowledges the support in part from Shuimu fellowships at Tsinghua University and the EPSRC under grant EP/X030881/1.

\bibliographystyle{apsreve}
\let\oldaddcontentsline\addcontentsline
\renewcommand{\addcontentsline}[3]{}
\bibliography{ref.bib}
\let\addcontentsline\oldaddcontentsline
\onecolumngrid

\clearpage
\newpage
\widetext

\begin{center}
\textbf{\large Supplemental Material for ``Supersymmetry dynamics on Rydberg atom arrays''}
\end{center}

\renewcommand{\thefigure}{S\arabic{figure}}
\setcounter{figure}{0}
\renewcommand{\theequation}{S\arabic{equation}}
\setcounter{equation}{0}
\renewcommand{\thesection}{\Roman{section}}
\setcounter{section}{0}
\setcounter{secnumdepth}{4}

\addtocontents{toc}{\protect\setcounter{tocdepth}{0}}
{
\tableofcontents
}

\section{The advantage of demonstrating supersymmetry via supercharge dynamics}
Besides the constant supercharge expectation value, other quantum dynamics features can also demonstrate supersymmetry. As previously investigated in Ref.~\cite{PhysRevLett.128.050504}, if the Hamiltonian $H$ is supersymmetric, the dynamics of the fidelity between the right-most one-kink state and the evolved left-most one-kink state
\begin{eqnarray}
    F(t) = \vert \langle K_{r} \vert e^{-iHt} \vert K_{l} \rangle \vert^2,
\end{eqnarray}
is identical to the dynamics of the fidelity between the right-most one-skink state and the evolved left-most one-skink state,
\begin{eqnarray}
    \Tilde{F}(t) = \vert \langle \Tilde{K}_{r} \vert e^{-iHt} \vert \Tilde{K}_{l} \rangle \vert^2,
\end{eqnarray}
where $\vert \Tilde{K}_{l (r)}\rangle$ is the superpartner of $\vert K_{l(r)} \rangle$. For a normalized eigenstate $\vert s \rangle$ which satisfies $H \vert s \rangle = E_{s} \vert s \rangle$ and $Q^{\dagger} \vert s \rangle = 0$, it is easy to find its superpartner $\vert \Tilde{s} \rangle  = Q \vert s \rangle$. Obviously, $H\vert \Tilde{s} \rangle = E_{s} \vert \Tilde{s} \rangle $ and $Q \vert \Tilde{s} \rangle =0$, i.e., $\vert s \rangle$ and $\vert \Tilde{s} \rangle$ form a doublet. However, $\vert \Tilde{s} \rangle$ is unnormalized because of
\begin{eqnarray}
    \langle \Tilde{s} \vert \Tilde{s} \rangle =  \langle s \vert Q^{\dagger}Q \vert s \rangle = \langle s \vert H \vert s \rangle = E_{s},
\end{eqnarray}
which hinders the preparation of the superpartner of a generic state that is not an energy eigenstate.

We use $\vert \Tilde{s}^{\prime} \rangle = \vert  \Tilde{s} \rangle / \sqrt{E_{s}}$ to denote the normalized superpartner of $\vert s \rangle$. In general, the one-kink state or other easily prepared state $\vert \psi \rangle$ is a superposition of various eigenstates $\vert s_{i} \rangle$, i.e., $\vert \psi \rangle = \sum_{i} \alpha_{i} \vert s_{i} \rangle$. For simplicity, we assume all eigenstates $\vert s_{i} \rangle$ are normalized and satisfy $Q^{\dagger} \vert s_{i} \rangle = 0$. The eigenenergy of $\vert s_{i} \rangle$ is $H \vert s_{i} \rangle = E_{s_{i}} \vert s_{i} \rangle$. The superpartner of $\vert \psi \rangle$ is 
\begin{eqnarray}
    \vert \Tilde{\psi} \rangle = \sum_{i} \alpha_{i} \vert \Tilde{s}^{\prime}_{i} \rangle = \sum_{i} \frac{\alpha_{i}}{\sqrt{E_{s_{i}}}} \vert \Tilde{s}_{i} \rangle = \sum_{i}  \frac{\alpha_{i}}{\sqrt{E_{s_{i}}}} Q \vert s_{i} \rangle = Q \sum_{i}  \frac{\alpha_{i}}{\sqrt{E_{s_{i}}}} \vert s_{i} \rangle \neq \frac{Q}{z} \vert \psi \rangle,
\end{eqnarray}
where $z$ is the normalization factor $z = \sqrt{\langle \psi \vert Q^{\dagger} Q \vert \psi \rangle} = \sqrt{\sum_{i} \vert \alpha_{i} \vert^2 E_{s_{i}}}$. Consequently, we can not straightforwardly apply supercharge operator $Q$ or $Q^{\dagger}$ to a state to obtain its superpartner. Therefore, the state preparation is challenging for experimentally demonstrating the supersymmetry via the identical fidelity dynamics in different parity sectors. For the one-kink and the corresponding one-skink states considered in Ref.~\cite{PhysRevLett.128.050504}, although they can be prepared approximately, an adiabatic evolution with prior knowledge is necessary. In contrast, no high-precision initial state preparation is required for demonstrating supersymmetry dynamics via the expectation value of supercharge and thus our scheme is more friendly for experimental investigation. Please see the subsequent section for a detailed comparison between our work and other experimental proposals for SUSY on Rydberg atom arrays.

\section{Comparison with other experimental proposals for SUSY on Rydberg atom arrays}

In this section, we provide a detailed comparison between the proposal investigated in this work and other experimental proposals for SUSY on Rydberg atom arrays.

As mentioned above, Ref.~\cite{PhysRevLett.128.050504} proposed exploring the supersymmetry dynamics on Rydberg atom arrays via the identical fidelity dynamics in even and odd parity sectors. Our proposal for investigating supersymmetry dynamics offers several advantages over the one in Ref.~\cite{PhysRevLett.128.050504} for current experimental platforms, particularly in terms of initial state preparation, observable measurement, and experimental setup. 1. Easier initial state preparation. As detailed above, preparing the superpartner of a generic state is inherently challenging. While the specific skink and kink states, which are superpartners, can be prepared in the previous proposal in principle, the task requires an adiabatic evolution. In contrast, our proposal necessitates only pure time-independent Hamiltonian evolution on Rydberg atom arrays for initial state preparation. 2. Easier observable measurement. The former proposal necessitates measuring fidelity to demonstrate supersymmetry dynamics. Our proposal, however, requires only the measurement of Pauli strings, which is much easier than global quantities such as fidelity as further elaborated in the subsequent section. 3. Easier experimental setup. The previous proposal requires the integration of an optical lattice with Rydberg atom manipulation to engineer the supersymmetric Hamiltonian. Our proposal, on the other hand, only requires Floquet engineering, which is a more accessible approach for current experimental setups.

In addition, previous work~\cite{li2024uncovering} has demonstrated that the tricritical point of the Ising transition can be achieved on Rydberg atom arrays and proposed to explore the emergent spacetime SUSY on Rydberg atom arrays. Compared to this proposal, our work is also more favorable. 1. No two-qubit digital gate is required in observable measurement in our work while it is inevitable to measure the non-local string operators in the former proposal. 2. Although fine-tuning two parameters is necessary for both proposals, the tuned parameters are analytically trackable in our work. 3. The SUSY characteristics can be revealed solely from dynamics in our case. Although more intriguing spacetime SUSY can be evaluated on ground state, the ground state preparation task is very experimentally challenging, as it requires careful design on adiabatic evolution. Similarly, another work~\cite{sable2024quantumanaloguesimulationmanybody} proposed a normalized Witten index as an observable for SUSY and its breaking in supersymmetric lattice models, which can be estimated on Rydberg atom arrays. However, sophisticated engineered thermalization~\cite{Schönleber_2018, PhysRevResearch.2.023214} is required which also challenges the experimental demonstration.

\section{Initial state preparation and supercharge measurement}
In this section, we introduce how to prepare the initial state for supersymmetry dynamics and measure the expectation value of supercharge experimentally.

Starting from a $\vert g \rangle^{\otimes L}$ state, we can perform Rydberg $\pi$ pulses on the desired sites to create any product states consistent with blockade constraint, e.g., $\{3j \vert j \in[1, l] \}$ for $\vert K_{1} \rangle$ state considered in the main text. Then we can evolve $\vert K_{1} \rangle$ by pure PXP Hamiltonian to generate a superposition state in even and odd parity sectors with non-zero supercharge as the initial state for supersymmetry dynamics. Therefore, the initial state preparation is suitable for the current Rydberg atom array platform.

The supercharge operator shown in Eq.~\eqref{eq:supercharge} can be rewritten as
\begin{eqnarray}
    \label{eq:supercharge_spin}
    Q &=& (-1) c_{1}^{\dagger} P_{2} + \sum_{i=2}^{L-1} (-1)^{i} P_{i-1}c_{i}^{\dagger}P_{i+1} + (-1)^{L} P_{L-1} c_{L}^{\dagger} \\ \nonumber
    &=& d_{1}n^{d}_{2} + \sum_{i=2}^{L-1} n^{d}_{i-1} d_{i} n^{d}_{i+1} + n_{L-1}^{d} d_{L} \\ \nonumber
    &=& \frac{\sigma^{-}_{1}}{2} \frac{\mathbb{I}+\sigma^{z}_{2}}{2} + \sum_{i=2}^{L-1} \frac{\mathbb{I}+\sigma^{z}_{i-1}}{2} \left( \frac{\sigma^{-}_{i}}{2} \prod_{j=1}^{i-1} (-\sigma^{z}_{j}) \right) \frac{\mathbb{I}+\sigma^{z}_{i+1}}{2} + \frac{\mathbb{I}+\sigma^{z}_{L-1}}{2} \left(\frac{\sigma^{-}_{L}}{2} \prod_{j=1}^{L-1} (-\sigma_{j}^{z}) \right),
\end{eqnarray}
where $d_{i} \equiv (-1)^{i} c_{i}^{\dagger}$ and $n_{i}^{d} = d_{i}^{\dagger}d_{i} = P_{i}$, corresponding to the particle-hole transformation. Therefore, the expectation value of the supercharge can be obtained from the summation of the expectation values of a series of Pauli strings. We can employ the quantum Analog-to-Digital Converter (qADC)~\cite{bluvsteinQuantumProcessorBased2022a, RevModPhys.89.015006} that transforms a state in the analogy basis to a digital basis to measure the expectation value of a Pauli string. The analogy basis of Rydberg atom arrays is $\{ \vert g \rangle, \vert r \rangle \}$ and the digital basis is $\{ \vert 0 \rangle, \vert 1 \rangle \}$ which can correspond to two hyperfine spin levels of the ground state ($\vert 1 \rangle$ is identified as $\vert g \rangle$). As detailed in Ref.~\cite{bluvsteinQuantumProcessorBased2022a}, the qADC can be implemented through a coherent mapping protocol. Specifically, we can apply a Raman $\pi$ pulse to map $\vert g \rangle $ to $\vert 0 \rangle$ and then a subsequent Rydberg $\pi$ pulse to map $\vert r \rangle $ to $\vert 1 \rangle$. Once the state has been converted to the digital basis, the measurement of any Pauli string is easy via proper single-qubit rotations and computational basis measurements.

\section{$M_{1}$ model with open boundary conditions}
In this section, we present the derivation of the Hamiltonian of the $M_{1}$ model~\cite{PhysRevLett.90.120402} in the spin-1/2 basis with open boundary conditions. The supercharge operator $Q$ is
\begin{eqnarray}
    \label{eq:supercharge}
    Q = \sum_{i=1}^{L} Q_{i} = (-1) c_{1}^{\dagger} P_{2} + \sum_{i=2}^{L-1} (-1)^{i} P_{i-1}c_{i}^{\dagger}P_{i+1} + (-1)^{L} P_{L-1} c_{L}^{\dagger},
\end{eqnarray}
where $L$ is the system size, $c^{\dagger}_{i} \ (c_{i})$ is the creation (annihilation) operator on $i$-th site, and $P_{i} \equiv 1 - c^{\dagger}_{i}c_{i}$ is the projector. Therefore, the Hamiltonian of the $M_{1}$ model with open boundary conditions is
\begin{eqnarray}
    H^{\text{obc}}_{M_{1}} &=& \{Q, Q^{\dagger} \} \\ \nonumber
    &=& \{ \sum_{i}Q_{i}, \sum_{j} Q_{j}^{\dagger} \} \\ \nonumber
    &=& \sum_{i=1}^{L} \{Q_{i}, Q_{i}^{\dagger} \} + \sum_{i=1}^{L-1} \{Q_{i}, Q_{i+1}^{\dagger} \} + \sum_{i=2}^{L} \{Q_{i}, Q_{i-1}^{\dagger} \}  \\ \nonumber
    &=& P_{2} + \sum_{i=2}^{L-1} P_{i-1}P_{i+1} + P_{L-1} \\ \nonumber
    &-& c_{1}^{\dagger} c_{2} P_{3} - \sum_{i=2}^{L-2} P_{i-1}c_{i}^{\dagger} c_{i+1} P_{i+2} - P_{L-2} c_{L-1}^{\dagger} c_{L} \\ \nonumber
    &-& c_{2}^{\dagger} c_{1} P_{3} - \sum_{i=2}^{L-2} P_{i-1}c_{i+1}^{\dagger} c_{i} P_{i+2} - P_{L-2} c_{L}^{\dagger} c_{L-1}.
\end{eqnarray}
We note that the configuration with adjacent fermions is forbidden and the Hilbert space is constrained so that nearest-neighbor sites cannot be simultaneously occupied~\cite{PaulFendley_2003}, i.e.,
\begin{eqnarray}
    \label{eq:constraint_fermion}
    n_{i}n_{i+1} = 0. 
\end{eqnarray}
And thus
\begin{eqnarray}
    H^{\text{obc}}_{M_{1}} &=& n_{2}^{d} + \sum_{i=2}^{L-1} n_{i-1}^{d}n_{i+1}^{d} + n_{L-1}^{d} \\ \nonumber
    &-& (d_{1}^{\dagger} d_{2} + d_{2}^{\dagger}d_{1}) n^{d}_{3} - \sum_{i=2}^{L-2} n^{d}_{i-1} (d^{\dagger}_{i}d_{i+1}+d^{\dagger}_{i+1}d_{i}) n^{d}_{i+2} - n^{d}_{L-2}(d_{L-1}^{\dagger} d_{L} + d_{L}^{\dagger}d_{L-1}).
\end{eqnarray}
We can utilize the Jordan-Wigner transformation to map fermionic creation and annihilation operators onto spin operators,
\begin{eqnarray}
    \sigma_{i}^{z}  & \equiv& 2n_{i}^{d} - \mathbb{I} , \\ \nonumber
 \frac{1}{2}(\sigma_{i}^{x}\sigma_{i+1}^{x}+\sigma^{y}_{i} \sigma_{i+1}^{y}) &\equiv& (d_{i}^{\dagger}d_{i+1}+d_{i+1}^{\dagger}d_{i}).
\end{eqnarray}
We define the projection operators on the spin basis as
\begin{eqnarray}
    \label{eq:projection}
    P_{i}^{r} &=& \frac{\mathbb{I}+\sigma^{z}_{i}}{2}, \\ \nonumber
    n^{r}_{i} &=& \frac{\mathbb{I}- \sigma^{z}_{i}}{2},
\end{eqnarray}
which projects site $i$ onto ground and Rydberg states respectively. Consequently, 
\begin{eqnarray}
    \label{eq:HM1}
    H^{\text{obc}}_{M_{1}} &=& P_{2}^{r}+\sum_{i=2}^{L-1} P^{r}_{i-1} P_{i+1}^{r} + P_{L-1}^{r}  \\ \nonumber
    &-& \frac{1}{2}(\sigma^{x}_{1}\sigma^{x}_{2}+\sigma^{y}_{1}\sigma^{y}_{2})P_{3}^{r} - \sum_{i=2}^{L-2} P_{i-1}^{r}\frac{1}{2}(\sigma^{x}_{i}\sigma^{x}_{i+1} + \sigma^{y}_{i}\sigma^{y}_{i+1}) P_{i+2}^{r} - P_{L-2}^{r}\frac{1}{2}(\sigma^{x}_{L-1}\sigma^{x}_{L}+\sigma^{y}_{L-1}\sigma^{y}_{L}) \\ \nonumber
    &=& \frac{1}{2} (\mathbb{I} + \sigma^{z}_{2}) + \sum_{i=2}^{L-1} \frac{1}{4}(\mathbb{I} + \sigma^{z}_{i-1} + \sigma^{z}_{i+1} + \sigma_{i-1}^{z}\sigma_{i+1}^{z}) + \frac{1}{2} (\mathbb{I} + \sigma^{z}_{L-1}) - H^{\text{obc}}_{\text{PXYP}} \\ \nonumber
    &=& \frac{1}{4 }\sum_{i=2}^{L-1} \sigma^{z}_{i-1} \sigma^{z}_{i+1} - H^{\text{obc}}_{\text{PXYP}} + \frac{1}{2}(\sigma^{z}_{2}+\sigma^{z}_{L-1}) + \frac{1}{4} \sum_{i=2}^{L-1} (\sigma^{z}_{i-1} + \sigma^{z}_{i+1}) + \text{const} \\ \nonumber
    &=& \frac{1}{4} H^{\text{obc}}_{\text{ZIZ}} - H^{\text{obc}}_{\text{PXYP}} - (n^{r}_{2}+n^{r}_{L-1}) -\frac{1}{2} \sum_{i=2}^{L-1} (n^{r}_{i-1}+n^{r}_{i+1}) + \text{const} \\ \nonumber
    &=& \frac{1}{4} H^{\text{obc}}_{\text{ZIZ}} - H^{\text{obc}}_{\text{PXYP}} - N + \frac{1}{2}(n^{r}_{1}-n^{r}_{2}-n^{r}_{N-1} + n^{r}_{N}) + \text{const}.
\end{eqnarray}
The aforementioned constraint Eq.~\eqref{eq:constraint_fermion} corresponds to
\begin{eqnarray}
    \label{eq:constraint}
    n_{i}n_{i+1} &=& 0 \\ \nonumber
    \Leftrightarrow (1-n^{d}_{i})(1-n^{d}_{i+1}) &=& 0 \\ \nonumber
    \Leftrightarrow n^{r}_{i}n^{r}_{i+1} &=& 0,
\end{eqnarray}
i.e., the Rydberg blockade.

\section{Derivation of the effective Floquet Hamiltonian}
In this section, we present the detailed derivation of the effective Floquet Hamiltonian $H_{F}$ with open boundary conditions. For more details, please refer to Ref.~\cite{Floquet}.

\subsection{Hamiltonian engineering}
The Hamiltonian of the driven PXP model with open boundary conditions is
\begin{eqnarray}
    \label{eq:H0}
    H^{\text{obc}}_{0}(t) = \frac{\Omega}{2}H^{\text{obc}}_{\text{PXP}}-\Delta_{0}(t) N,
\end{eqnarray}
where 
\begin{eqnarray}
    H^{\text{obc}}_{\text{PXP}} &=& \sigma^{x}_{1}P^{r}_{2} + \sum_{i=2}^{L-1} P^{r}_{i-1}\sigma^{x}_{i} P^{r}_{i+1} + P^{r}_{L-1} \sigma^{x}_{L}, \\ 
    N &=& \sum_{i}n^{r}_{i}, 
\end{eqnarray}
and $\Delta_{0}(t)$ is the time-dependent global detuning given by
\begin{eqnarray}
    \label{eq:delta0}
    \Delta_{0}(t) &=& \pi \sum_{n} \delta(t-\frac{\tau}{4}-n\frac{\tau}{2}),
\end{eqnarray}
with $\tau$ being the Floquet period. It generates a Floquet unitary with period $\tau/2$
\begin{eqnarray}
    \chi_{\tau/2} = U_{0}(\tau/2) &=& e^{-i \frac{\tau}{4} \frac{\Omega}{2} H^{\text{obc}}_{\text{PXP}}} e^{i\pi N} e^{-i \frac{\tau}{4} \frac{\Omega}{2} H^{\text{obc}}_{\text{PXP}}} \\ \nonumber
    &=& e^{i\pi N} e^{+i \frac{\tau}{4} \frac{\Omega}{2} H^{\text{obc}}_{\text{PXP}}}  e^{-i \frac{\tau}{4} \frac{\Omega}{2} H^{\text{obc}}_{\text{PXP}}} \\ \nonumber
    &=& e^{i\pi N},
\end{eqnarray}
and thus 
\begin{eqnarray}
    \label{eq:U0tau}
    U_{0}(\tau) = \mathbb{I},
\end{eqnarray}
i.e., the time-dependent detuning (Eq.~\eqref{eq:delta0}) realizes a many-body echo at stroboscopic times $n\tau$. Between stroboscopic times, $0 \leq t \ < \tau$,
\begin{eqnarray}
    U_{0}(t) = e^{-i \pi N \cdot \mathbb{I}_{\tau/4 \leq t < 3\tau/4}} e^{-it^{\prime} \frac{\Omega}{2} H_{\text{PXP}}^{\text{obc}}},
\end{eqnarray}
where $t^{\prime} = t^{\prime}(t) = ||t-\tau/4 | - \tau/2 | - \tau/4$~\cite{Floquet}.

We consider perturbations around the many-body echo point $H_{0}(t)$, in the form of additional detuning pulses coupling to the number operator $N$ 
\begin{eqnarray}
    H(t) = H^{\text{obc}}_{0}(t) - \Tilde{\Delta}(t) N,
\end{eqnarray}
where 
\begin{eqnarray}
    \label{eq:deltatilde}
    \Tilde{\Delta}(t) &=& (\gamma-\theta) \sum_{n} \delta(t-n\tau) + \theta \sum_{n} \delta(t-\frac{\tau}{2}-n\tau) + \epsilon \sum_{n} \delta(t-\frac{\tau}{4}-n\tau) + \epsilon \sum_{n} \delta(t-\frac{3\tau}{4}-n\tau) \\ \nonumber
    &=& \sum_{j,n} \Tilde{\Delta}_{j} \delta(t-t_j-n\tau).
\end{eqnarray}
The associated dynamics can be analyzed in a frame co-rotating with $H_{0}(t)$,
\begin{eqnarray}
    \Tilde{H}(t) = -\Tilde{\Delta}(t) U_{0}(t)^{\dagger} N U_{0}(t),
\end{eqnarray}
and the dynamics coincide with those in the laboratory frame at times $n\tau$ due to Eq.~\eqref{eq:U0tau}. The number operator in this rotated frame is
\begin{eqnarray}
    \label{eq:number_rotated}
    \Tilde{N}(t^{\prime}) &=& U_{0}(t)^{\dagger} N U_{0}(t)  \\ \nonumber
    &=& e^{it^{\prime} \frac{\Omega}{2} H^{\text{obc}}_{\text{PXP}}} N e^{-it^{\prime}\frac{\Omega}{2}H^{\text{obc}}_{\text{PXP}}} \\ \nonumber
    &=& N + \frac{\Omega}{2}it^{\prime} [H^{\text{obc}}_{\text{PXP}}, N] - \frac{\Omega^2}{8}(t^{\prime})^2 [H^{\text{obc}}_{\text{PXP}},[H^{\text{obc}}_{\text{PXP}}, N]] + \dots,
\end{eqnarray}
and thus
\begin{eqnarray}
    \Tilde{H}(t) &=& -\sum_{j,n} \Tilde{\Delta}_{j} \delta(t-t_{j}-n\tau) \Tilde{N}(t_{j}^{\prime}) \\ \nonumber
    &=& - \left((\gamma-\theta) \sum_{n} \delta(t-n\tau) \Tilde{N}(0) + \theta \sum_{n} \delta(t-\frac{\tau}{2}-n\tau) \Tilde{N}(0)+ \epsilon \sum_{n} \delta(t-\frac{\tau}{4}-n\tau) \Tilde{N}(\frac{\tau}{4}) + \epsilon \sum_{n} \delta(t-\frac{3\tau}{4}-n\tau)\Tilde{N}(-\frac{\tau}{4}) \right),
\end{eqnarray}
i.e., the pulses at $t_{j} = (0, \frac{\tau}{2})$ couple to the bare Rydberg number operator $\Tilde{N}(0) = N$, while pulses at $t_{j} = (\frac{\tau}{4}, \frac{3\tau}{4})$ couple to $\Tilde{N}(\pm \frac{\tau}{4})$.
The stroboscopic dynamics can be described by the Floquet unitary $U_{F}$, and 
\begin{eqnarray}
    U(n\tau) &=& (U_{F})^{n}, \\ 
    U_{F} &=& e^{i\epsilon \Tilde{N}(-\frac{\tau}{4})} e^{i \theta \Tilde{N}(0)} e^{i\epsilon \Tilde{N}(\frac{\tau}{4})}e^{i(\gamma-\theta)\Tilde{N}(0)}.
\end{eqnarray}

\subsection{Number operator in the rotated frame}
As discussed above, the stroboscopic dynamics can be described by the Floquet unitary $U_{F}$ which can be generated by the global number operator in the rotated frame $\Tilde{N}(t^{\prime})$. We now introduce the derivation of $\Tilde{N}(t^{\prime})$.

In a regime of small Floquet periods $\frac{\Omega \tau}{4} \ll 1$, the number operator shown in Eq.~\eqref{eq:number_rotated} is 
\begin{eqnarray}
    \label{eq:number_rotated_pertuebation}
    \Tilde{N}(t^{\prime}) &=& N + \frac{\Omega}{2}it^{\prime} [H^{\text{obc}}_{\text{PXP}}, N] - \frac{\Omega^2}{8}(t^{\prime})^2 [H^{\text{obc}}_{\text{PXP}},[H^{\text{obc}}_{\text{PXP}}, N]] + \mathcal{O}(\Omega^3 \tau^3).
\end{eqnarray}
We have
\begin{eqnarray}
    [H^{\text{obc}}_{\text{PXP}}, N] &=& [H^{\text{obc}}_{\text{PXP}}, \sum_{j} n_{j}] \\ \nonumber
    &=& -\frac{1}{2} \sum_{i} [H^{\text{obc}}_{\text{PXP}}, \sigma^{z}_{i}] \\ \nonumber
    &=& -\frac{1}{2} \left([\sigma^{x}_{1}P^{r}_{2}, \sigma^{z}_{1}] + \sum_{j=2}^{L-1} [P^{r}_{j-1}\sigma^{x}_{j}P^{r}_{j+1}, \sigma^{z}_{j}] + [P^{r}_{L-1}\sigma^{x}_{L}, \sigma^{z}_{L}] \right) \\ \nonumber
    &=& i H^{\text{obc}}_{\text{PYP}},
\end{eqnarray}
and
\begin{eqnarray}
    \label{eq:second_order}
    [H^{\text{obc}}_{\text{PXP}}, [H^{\text{obc}}_{\text{PXP}}, N]] &=& i [H^{\text{obc}}_{\text{PXP}}, H^{\text{obc}}_{\text{PYP}}] \\ \nonumber
    &=& i[\sigma^{x}_{1}, \sigma^{y}_{1}]P^{r}_{2} + i \sum_{j=2}^{L-1} P^{r}_{j-1}[\sigma^{x}_{j}, \sigma^{y}_{j}] P^{r}_{j+1}  + iP^{r}_{L-1}[\sigma^{x}_{L}, \sigma^{y}_{L}]\\ \nonumber
    &+& i [\sigma_{1}^{x}P^{r}_{2}, P^{r}_{1}\sigma^{y}_{2}] P^{r}_{3} + 
    i \sum_{j=2}^{L-2} P^{r}_{j-1} [\sigma_{j}^{x}P^{r}_{j+1}, P^{r}_{j}\sigma^{y}_{j+1}] P^{r}_{j+2}  + i P^{r}_{L-2}[\sigma_{L-1}^{x}P^{r}_{L}, P^{r}_{L-1}\sigma^{y}_{L}] \\ \nonumber
    &+& i [P^{r}_{1}\sigma^{x}_{2}, \sigma^{y}_{1}P^{r}_{2}]P^{r}_{3} + i\sum_{j=3}^{L-1} P^{r}_{j-2}[P^{r}_{j-1}\sigma^{x}_{j}, \sigma^{y}_{j-1}P^{r}_{j}] P^{r}_{j+1} + iP^{r}_{L-2} [P^{r}_{L-1}\sigma^{x}_{L}, \sigma^{y}_{L-1}P^{r}_{L}].
\end{eqnarray}
Because 
\begin{eqnarray}
    [\sigma^{x}_{j}, \sigma^{y}_{j}] &=& 2i \sigma^{z}_{j}, \\ \nonumber
    [\sigma^{x}_{j-1}P^{r}_{j}, P^{r}_{j-1}\sigma^{y}_{j}] &=& -\frac{i}{2} (\sigma_{j-1}^{x}\sigma_{j}^{x}+\sigma_{j-1}^{y}\sigma_{j}^{y}),\\ \nonumber
   [P^{r}_{j-1}\sigma^{x}_{j}, \sigma^{y}_{j-1}P^{r}_{j}]   &=&  -\frac{i}{2} (\sigma_{j-1}^{x}\sigma_{j}^{x}+\sigma_{j-1}^{y}\sigma_{j}^{y}),
\end{eqnarray}
and thus Eq.~\eqref{eq:second_order} is
\begin{eqnarray}
     [H^{\text{obc}}_{\text{PXP}}, [H^{\text{obc}}_{\text{PXP}}, N]] &=& -2(\sigma_{1}^{z}P^{r}_{2} + \sum_{j=2}^{L-1} P^{r}_{j-1} \sigma^{z}_{j}P^{r}_{j+1} + P^{r}_{L-1} \sigma^{z}_{L}) \\ \nonumber
     &+& (\sigma^{x}_{1}\sigma^{x}_{2} + \sigma^{y}_{1}\sigma^{y}_{2})P^{r}_{3} + \sum_{j=2}^{L-2} P^{r}_{j-1} (\sigma^{x}_{j}\sigma^{x}_{j+1} + \sigma^{y}_{j}\sigma^{y}_{j+1})P^{r}_{j+2} + P^{r}_{L-2}(\sigma^{x}_{L-1}\sigma^{x}_{L} + \sigma^{y}_{L-1}\sigma^{y}_{L}) \\ \nonumber
     &=& -2 H^{\text{obc}}_{\text{PZP}} + 2 H^{\text{obc}}_{\text{PXYP}}.
\end{eqnarray}
Therefore, the number operator in the rotated frame shown in Eq.~\eqref{eq:number_rotated_pertuebation} is
\begin{eqnarray}
    \Tilde{N}(t^{\prime}) \approx N - \frac{\Omega}{2} t^{\prime} H^{\text{obc}}_{\text{PYP}} + \frac{\Omega^2}{4}(t^{\prime})^{2} (H^{\text{obc}}_{\text{PZP}} - H^{\text{obc}}_{\text{PXYP}}).
\end{eqnarray}

\subsection{Effective Floquet Hamiltonian}
For small perturbations, the Floquet unitary $U_{F}$ is well-described by a static effective Floquet Hamiltonian,
\begin{eqnarray}
    U_{F} \approx e^{-i H_{F} \tau}.
\end{eqnarray}
The leading order contributions to the effective Floquet Hamiltonian can be obtained by a Floquet-Magnus expansion with
\begin{eqnarray}
    H_{F}^{(0)} &=& \sum_{j} -\frac{\Tilde{\Delta}_{j}}{\tau} \Tilde{N}(t^{\prime}_{j}), \\ \nonumber
    H_{F}^{(1)} &=& \sum_{j > l} \frac{\Tilde{\Delta}_{j}\Tilde{\Delta}_l}{2i\tau} [\Tilde{N}(t^{\prime}_{j}), \Tilde{N}(t^{\prime}_{l})].
\end{eqnarray}
For the specific parameterization of detuning perturbations as shown in Eq.~\eqref{eq:deltatilde}, the associated effective Floquet Hamiltonian is 
\begin{eqnarray}
    \label{eq:Heff}
    \tau H_{F}^{(0)} &=& -\gamma \Tilde{N}(0) - \epsilon (\Tilde{N}(\frac{\tau}{4}) + \Tilde{N}(-\frac{\tau}{4})), \\ \nonumber
    \Rightarrow H_{F}^{(0)} &=& -\frac{\gamma}{\tau} N - \frac{\epsilon}{\tau} (2N+\frac{\Omega^2 \tau^2}{32} (H^{\text{obc}}_{\text{PZP}}-H^{\text{obc}}_{\text{PXYP}})), \\ \nonumber
    &=& -\frac{(\gamma+2\epsilon)}{\tau} N - \frac{\epsilon \Omega^2 \tau}{32} (H^{\text{obc}}_{\text{PZP}}-H^{\text{obc}}_{\text{PXYP}}),
\end{eqnarray}
and
\begin{eqnarray}
    2i\tau H_{F}^{(1)} &=& \epsilon(\gamma-2\theta)[\Tilde{N}(0), \Tilde{N}(-\frac{\tau}{4})] + \epsilon \gamma [\Tilde{N}(0), \Tilde{N}(\frac{\tau}{4})] + \epsilon^2 [\Tilde{N}(-\frac{\tau}{4}), \Tilde{N}(\frac{\tau}{4})]. 
\end{eqnarray}
We have
\begin{eqnarray}
    \label{eq:hf1_1}
    [\Tilde{N}(0), \Tilde{N}(-\frac{\tau}{4})] &=& [N, N+\frac{\Omega \tau}{8} H^{\text{obc}}_{\text{PYP}}+\frac{\Omega^2 \tau^2}{64}(H^{\text{obc}}_{\text{PZP}}-H^{\text{obc}}_{\text{PXYP}})], \\ \nonumber
    &=& [N, \frac{\Omega \tau}{8} H^{\text{obc}}_{\text{PYP}}-\frac{\Omega^2 \tau^2}{64}H^{\text{obc}}_{\text{PXYP}}] \\ \nonumber
    &=& i \frac{\Omega\tau}{8} H^{\text{obc}}_{\text{PXP}},
\end{eqnarray}
because 
\begin{eqnarray}
    [N, H^{\text{obc}}_{\text{PXYP}}] &=& 0 \\ \nonumber
    \Leftrightarrow [\sigma^{z}_{j}, (\sigma^{x}_{j}\sigma^{x}_{j+1}+\sigma^{y}_{j}\sigma^{y}_{j+1}) + (\sigma^{x}_{j-1}\sigma^{x}_{j}+\sigma^{y}_{j-1}\sigma^{y}_{j})] &=& 0.
\end{eqnarray}
Similarly,
\begin{eqnarray}
    [\Tilde{N}(0), \Tilde{N}(\frac{\tau}{4})] &=& [N, N-\frac{\Omega \tau}{8} H^{\text{obc}}_{\text{PYP}}+\frac{\Omega^2 \tau^2}{64}(H^{\text{obc}}_{\text{PZP}}-H^{\text{obc}}_{\text{PXYP}})], \\ \nonumber
    &=& [N, -\frac{\Omega \tau}{8} H^{\text{obc}}_{\text{PYP}}]  \\ \nonumber
    &=& -i \frac{\Omega\tau}{8} H^{\text{obc}}_{\text{PXP}}, 
\end{eqnarray}
\begin{eqnarray}
    \label{eq:hf1_3}
    [\Tilde{N}(-\frac{\tau}{4}), \Tilde{N}(\frac{\tau}{4})] &=& [N+\frac{\Omega \tau}{8} H^{\text{obc}}_{\text{PYP}}+\frac{\Omega^2 \tau^2}{64}(H^{\text{obc}}_{\text{PZP}}-H^{\text{obc}}_{\text{PXYP}}), N-\frac{\Omega \tau}{8} H^{\text{obc}}_{\text{PYP}}+\frac{\Omega^2 \tau^2}{64}(H^{\text{obc}}_{\text{PZP}}-H^{\text{obc}}_{\text{PXYP}})] \\ \nonumber
    &\approx& [N, -\frac{\Omega \tau}{8} H^{\text{obc}}_{\text{PYP}}] + [\frac{\Omega \tau}{8} H^{\text{obc}}_{\text{PYP}}, N] \\ \nonumber
    &=& -i \frac{\Omega\tau}{4} H^{\text{obc}}_{\text{PXP}}.
\end{eqnarray}

Consequently,
\begin{eqnarray}
    H_{F}^{(1)} &=& \frac{1}{2\tau} (\epsilon(\gamma-2\theta) -\epsilon \gamma - 2\epsilon^2) \frac{\Omega \tau}{8} H^{\text{obc}}_{\text{PXP}} \\ \nonumber
    &=&  (- \epsilon \theta  - \epsilon^2) \frac{\Omega}{8} H^{\text{obc}}_{\text{PXP}}.
\end{eqnarray}
The effective Floquet Hamiltonian is 
\begin{eqnarray}
    \label{eq:HF_PZP}
    H_{F} &=& H_{F}^{(0)} + H_{F}^{(1)} \\ \nonumber
    &=& -\frac{(\gamma+2 \epsilon)}{\tau} N - \frac{\epsilon \Omega^2 \tau}{32} ({H^{\text{obc}}_{\text{PZP}}}-H^{\text{obc}}_{\text{PXYP}}) - (\epsilon \theta + \epsilon^2) \frac{\Omega }{8} H^{\text{obc}}_{\text{PXP}}.
\end{eqnarray}
Moreover,
\begin{eqnarray}
    H^{\text{obc}}_{\text{PZP}} &=&  \sigma^{z}_{1} P^{r}_{2} + \sum_{i=2}^{L-1} P^{r}_{i-1}\sigma^{z}_{i} P^{r}_{i+1} + P^{r}_{L-1} \sigma^{z}_{L} \\ \nonumber
    &=& \sigma^{z}_{1} \frac{\mathbb{I} + \sigma^{z}_{2}}{2} + \sum_{i=2}^{L-1} \frac{\mathbb{I}+\sigma^{z}_{i-1}}{2}(1-2n^{r}_{i}) \frac{\mathbb{I}+\sigma^{z}_{i+1}}{2}  +  \frac{\mathbb{I} + \sigma^{z}_{L-1}}{2} \sigma^{z}_{L}\\ \nonumber
    &=& \frac{1}{4} H^{\text{obc}}_{\text{ZIZ}} + \frac{\sigma^{z}_{1}+\sigma^{z}_{L}}{2} + \frac{\sigma^{z}_{1}\sigma^{z}_{2}+\sigma^{z}_{L-1}\sigma^{z}_{L}}{2} + \frac{1}{4} \sum_{i=2}^{L-1}\left( 1 - 2n^{r}_{i} + \sigma^{z}_{i-1} + \sigma^{z}_{i+1} - 2 \sigma^{z}_{i-1}n^{r}_{i} - 2n^{r}_{i} \sigma^{z}_{i+1} - 2\sigma^{z}_{i-1} n^{r}_{i} \sigma^{z}_{i+1} \right)\\ \nonumber
    &=& \frac{1}{4} H^{\text{obc}}_{\text{ZIZ}} + \frac{(1-2n^{r}_{1})+(1-2n^{r}_{L})}{2} + \frac{(1-2n^{r}_{1})(1-2n^{r}_{2})+ (1-2n^{r}_{L-1})(1-2n^{r}_{L})}{2}  \\ \nonumber
    &-&\frac{1}{2} \sum_{i=2}^{L-1} n_{i}^{r} +\frac{1}{4} \sum_{i=2}^{L-1} (1-2n^{r}_{i-1} + 1-2n^{r}_{i+1}) -\frac{1}{2} \sum_{i=2}^{L-1}((1-2n^{r}_{i-1})n^{r}_{i} + n_{i}^{r} (1-2n^{r}_{i+1})) \\ \nonumber
    &-& \frac{1}{2} \sum_{i=2}^{L-1} (1-2n^{r}_{i-1}) n^{r}_{i} (1-2n^{r}_{i+1}) +\text{const} \\ \nonumber
    &=& \frac{1}{4} H^{\text{obc}}_{\text{ZIZ}} - 2n_{1}^{r} -n_{2}^{r}- n_{L-1}^{r} - 2n_{L}^{r}  + 2n_{1}^{r}n_{2}^{r} + 2n_{L-1}^{r}n_{L}^{r} - \frac{1}{2} \sum_{i=2}^{L-1} (4n_{i}^{r} + n_{i-1}^{r} + n_{i+1}^{r})  \\ \nonumber
    &+& \sum_{i=2}^{L-1} 2\left( n^{r}_{i-1}n_{i}^{r}+n^{r}_{i}n^{r}_{i+1} - n_{i-1}^{r}n^{r}_{i} n^{r}_{i+1} \right) + \text{const}.
\end{eqnarray}
As discussed above, $n_{i}^{r}n^{r}_{i+1}$ is zero in the constraint Hilbert space (Eq.~\eqref{eq:constraint}) and thus
\begin{eqnarray}
    H^{\text{obc}}_{\text{PZP}} = \frac{1}{4} H^{\text{obc}}_{\text{ZIZ}} -3N + \frac{1}{2} (n^{r}_{1}-n^{r}_{2}-n^{r}_{L-1}+n^{r}_{L}),
\end{eqnarray}
up to an unimportant constant.

Therefore, the effective Floquet Hamiltonian (see Eq.~\eqref{eq:HF_PZP}) is 
\begin{eqnarray}
 H_{F} &=& -\frac{(\gamma+2\epsilon)}{\tau} N - \frac{\epsilon \Omega^{2} \tau}{32} ((\frac{1}{4} H^{\text{obc}}_{\text{ZIZ}} -3N + \frac{1}{2} (n^{r}_{1}-n^{r}_{2}-n^{r}_{L-1}+n^{r}_{L}))-H^{\text{obc}}_{\text{PXYP}}) - \frac{\epsilon(\epsilon+\theta)\Omega}{8} H_{\text{PXP}}^{\text{obc}} \\ \nonumber
 &=& -(\frac{\gamma + 2\epsilon}{\tau} - \frac{3\epsilon \Omega^2 \tau}{32}) N - (-\frac{\epsilon \Omega^2 \tau}{32}) H^{\text{obc}}_{\text{PXYP}} + \frac{1}{4} (-\frac{\epsilon \Omega^2 \tau}{32})H^{\text{obc}}_{\text{ZIZ}} + \frac{1}{2} (-\frac{\epsilon \Omega^2 \tau}{32}) (n^{r}_{1}-n^{r}_{2}-n^{r}_{L-1}+n^{r}_{L})) \\ \nonumber 
&-& \frac{\epsilon(\epsilon+\theta)\Omega}{8} H_{\text{PXP}}^{\text{obc}}.
\end{eqnarray}
We set 
\begin{eqnarray}
    J &=& \frac{\gamma + 2\epsilon}{\tau} - \frac{3\epsilon \Omega^2 \tau}{32}, \\ \nonumber
    h &=& -\frac{\epsilon \Omega^2 \tau}{32}, \\ \nonumber
    g &=& -\frac{\epsilon(\epsilon+\theta)\Omega}{8}.
\end{eqnarray}
Then 
\begin{eqnarray}
    H_{F} &=& -J N -h H_{\text{PXYP}}^{\text{obc}} + \frac{h}{4} H^{\text{obc}}_{\text{ZIZ}} + \frac{h}{2}(n^{r}_{1}-n^{r}_{2}-n^{r}_{L-1}+n^{r}_{L})) + g H_{\text{PXP}}^{\text{obc}}.
\end{eqnarray}
When $\theta = -\epsilon$ and $\gamma = \epsilon (\frac{\Omega^2 \tau^2}{16} - 2)$, $g=0$, $J=h= - \frac{\epsilon \Omega^2 \tau}{32}$, and,
\begin{eqnarray}
    \label{eq:HF_J}
    H_{F} = -J N -J H^{\text{obc}}_{\text{PXYP}} + \frac{J}{4} H^{\text{obc}}_{\text{ZIZ}} + \frac{J}{2}(n^{r}_{1}-n^{r}_{2}-n^{r}_{N-1}+n^{r}_{N})),
\end{eqnarray}
which is the same as the Hamiltonain of the one-dimensional $M_{1}$ model with open boundary conditions shown in Eq.~\eqref{eq:HM1}.

\section{Details of numerical simulation and additional numerical results}
\begin{figure}[t]
\centering
\includegraphics[width=0.65\textwidth, keepaspectratio]{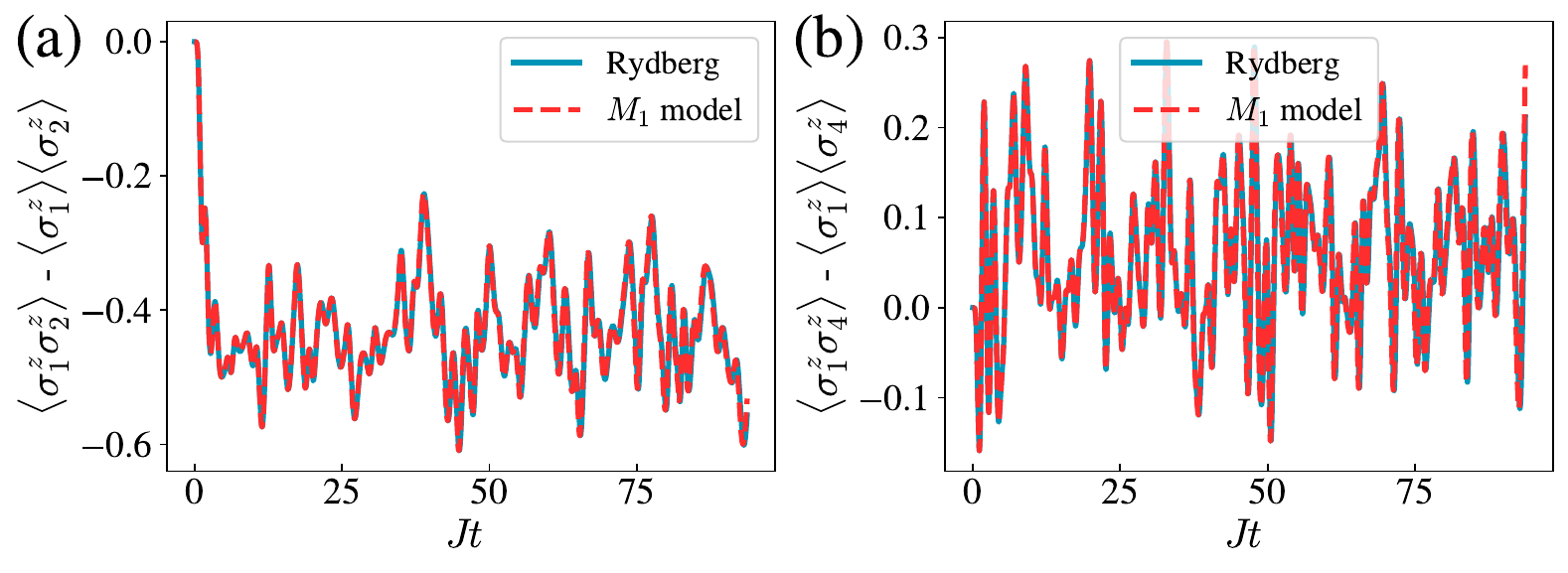}
\caption{The dynamics of $zz$ correlation: $\langle \sigma^{z}_{i}\sigma^{z}_{j}\rangle - \langle \sigma^{z}_{i}\rangle \langle \sigma^{z}_{j}\rangle$. (a) $i=1$ and $j=2$, (b) $i=1$ and $j=4$.
Here, we set $L=13$, $\tau=0.001$, $\frac{\Omega\tau}{4}=0.025$, $\epsilon=-0.1$. The numerical results evolved by the $M_{1}$ model (red) agree well with those of the Rydberg atom array (blue).}
\label{fig:zz_dynamics_l4_Omegatau0.1_epsilon0.1}
\end{figure}

\begin{figure}[t]
\centering
\includegraphics[width=0.65\textwidth, keepaspectratio]{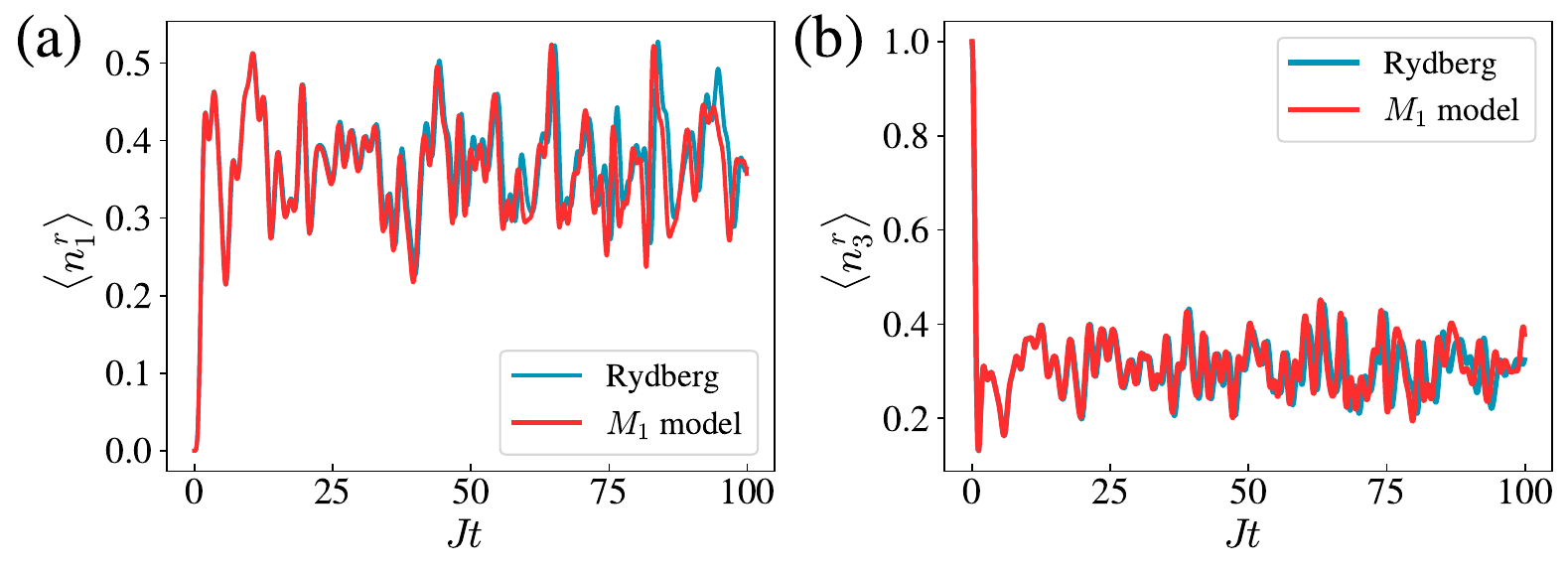}
\caption{The Rydberg state density dynamics on site 1 (a) and 3 (b) respectively. Here, we set $L=13$, $\tau=0.001$, $\frac{\Omega\tau}{4}=0.1$, $\epsilon=-0.2$. The numerical results evolved by the $M_{1}$ model (red) agree well with those of the Rydberg atom array (blue) in the early times.}
\label{fig:z_dynamics_l4_Omegatau0.4_epsilon0.2}
\end{figure}

\begin{figure}[t]
\centering
\includegraphics[width=0.65\textwidth, keepaspectratio]{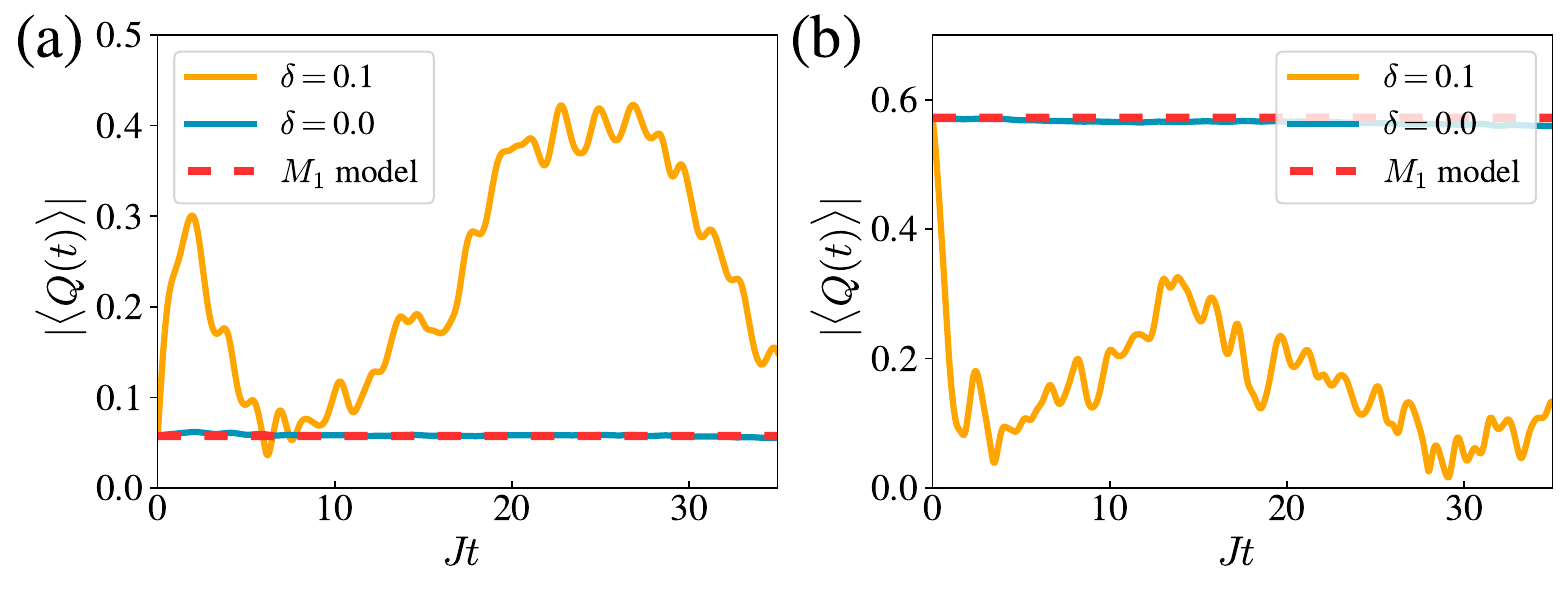}
\caption{Supercharge dynamics with (a) $t^{*}=5$ and (b) $t^{*}=10$. Here, we set $L=13$, $\tau=0.001$, $\frac{\Omega\tau}{4}=0.05$, $\epsilon=-0.1$, and $\theta = -(1+\delta) \epsilon$. The expectation value of supercharge is the same as that under the quench of $M_{1}$ model and is a constant when $\delta=0.0$. At the same time, the supercharger dynamics is nonconserved when $\delta\neq 0.0$, i.e., the effective Floquet Hamiltonian is nonsupersymmetric. We note that the accumulated approximation errors induce the deviation at late times even with $\delta=0$.}
\label{fig:Rydberg_Q_l4_Omegatau0.2_epsilon0.1}
\end{figure}

We employ the python package {\sf TensorCircuit}~\cite{*[{ }] [{https://github.com/tensorcircuit/tensorcircuit-ng}] Zhang2023tensorcircuit} to perform numerical simulations for Hamiltonian dynamics. As discussed above, we set 
\begin{eqnarray}
    \theta &=& - \epsilon, \\ \nonumber
    \gamma &=& \epsilon(\frac{\Omega^{2}\tau^2}{16}-2).
\end{eqnarray}
To demonstrate the correspondence between the stroboscopic dynamics of the driven PXP model under Floquet engineering and the supersymmetric $M_{1}$ model, we also calculated the $zz$ correlation $\langle \sigma^{z}_{i}\sigma^{z}_{j}\rangle - \langle \sigma^{z}_{i}\rangle \langle \sigma^{z}_{j}\rangle$. As shown in Fig~\ref{fig:zz_dynamics_l4_Omegatau0.1_epsilon0.1}, the results also coincide.
Moreover, we also compared the stroboscopic dynamics of the driven PXP model with larger $\frac{\Omega \tau}{4}$ and the dynamics of the $M_{1}$ model, as shown in Fig.~\ref{fig:z_dynamics_l4_Omegatau0.4_epsilon0.2} and Fig.~\ref{fig:Rydberg_Q_l4_Omegatau0.2_epsilon0.1}. The deviation appears much earlier than that shown in the main text due to the larger $\frac{\Omega \tau}{4}$, consistent with the theoretical prediction on the approximation error.

\end{document}